\documentclass[aps,prb,twocolumn,showpacs,superscriptaddress,amsmath,amssymb,floatfix,10pt,reprint]{revtex4-1}  
\usepackage{graphicx}  
\usepackage{dcolumn}   
\usepackage{bm}        
\usepackage{amssymb}   
\usepackage{framed}
\usepackage{amsmath}
\usepackage{hhline}
\usepackage{color}
\definecolor{bluegreen}{rgb}{0,0.2,0.8}
\usepackage{subfigure,amsmath,verbatim,moreverb}
\usepackage{tabularx}
\usepackage{lipsum}
\usepackage{longtable}
\usepackage{booktabs}
\usepackage{latexsym}
\usepackage{dcolumn}
\usepackage{amsmath}
\usepackage{epsf}
\usepackage{float}
\usepackage{hyperref}

\usepackage{etoolbox}
\AtBeginEnvironment{align}{\setcounter{subeqn}{0}}
\newcounter{subeqn} %

\begin{document}

\title{Accurate water properties from an efficient ab-initio method}
\author{Subrata Jana}
\email{subrata.jana@niser.ac.in}
\affiliation{School of Physical Sciences, National Institute of Science Education and Research, HBNI, 
Bhubaneswar 752050, India}
\author{Lucian A. Constantin}
\email{lucian.constantin@iit.it}
\affiliation{Center for Biomolecular Nanotechnologies @UNILE, Istituto Italiano di Tecnologia, 
Via Barsanti, I-73010 Arnesano, Italy}
\author{Prasanjit Samal}
\email{psamal@niser.ac.in}
\affiliation{School of Physical Sciences, National Institute of Science Education and Research, HBNI, Bhubaneswar 752050, India}

\date{\today}

\begin{abstract}

Accurate prediction of the water properties from a low-cost ab-initio method still a foremost problem for chemists and physicist. 
Though density functional approaches starting from semilocal to hybrid functionals are tested, those are not efficiently performed 
for all the properties together, especially, considering energies, conformal ranking, structural and dynamics of water. Also, the 
inclusion of the long-range van der Waals (vdW) interaction does not improve the ordering stability of isomer. However, relying on 
the simple revision of the Tao-Mo (revTM) semilocal meta-generalized gradient approximations, we demonstrate that all properties of 
the water can be accurately predicted. A consistent improvement over several popular ab-initio methods is achieved, indicating the 
accuracy of this method for describing hydrogen bonding of water.

\end{abstract}

\maketitle

\section{Introduction}

Density functional theory (DFT)~\cite{PhysRev.136.B864} is the {\it de-facto} standard method to calculate the 
electronic properties of molecules, solids, and liquids. DFT, which relies on the accuracy of the 
different levels of exchange-correlation (xc) approximations
\cite{PhysRevB.23.5048,PhysRevA.38.3098,PhysRevB.37.785,PhysRevB.46.6671,PhysRevLett.77.3865,
PhysRevB.82.113104,doi:10.1021/ct200510s,PhysRevB.72.085108,PhysRevB.73.235116,PhysRevB.84.233103,
doi:10.1063/1.2912068,PhysRevLett.100.136406,PhysRevB.91.041120,PhysRevB.79.075126,PhysRevB.84.045126,
doi:10.1063/1.4766324,PhysRevB.80.035125,doi:10.1021/ct200510s,PhysRevB.93.045126,doi:10.1063/1.5021597,
PhysRevLett.108.126402,PhysRevLett.106.186406,PhysRevA.39.3761,PhysRevB.88.125112,doi:10.1063/1.476577,
doi:10.1063/1.2370993,PhysRevLett.82.2544,PhysRevLett.91.146401,PhysRevLett.103.026403,PhysRevB.86.035130,
PhysRevLett.97.223002,doi:10.1021/ct400148r,doi:10.1063/1.4789414,Sun685,doi:10.1021/ct300269u,PhysRevB.93.115127,
PhysRevLett.115.036402,PhysRevLett.117.073001,Wang8487,doi:10.1021/acs.jctc.8b00072,doi:10.1002/qua.25224,
doi:10.1021/acs.jpca.9b02921,PhysRevB.100.045147,C9CP03356D}, 
performs remarkably for different interactions with or 
without weak dispersion  correction \cite{PhysRevB.79.085104,doi:10.1063/1.4948636,PhysRevB.95.035118,
doi:10.1063/1.2835596,PhysRevB.84.035117,PhysRevB.79.155107,doi:10.1063/1.5040786,doi:10.1063/1.5047863,
PatraE9188,PhysRevLett.111.106401,PhysRevX.6.041005,doi:10.1063/1.2370993,B907148B,Peverati20120476,
doi:10.1063/1.460205,doi:10.1063/1.1626543,doi:10.1021/ct0502763,doi:10.1021/ct100466k,doi:10.1021/ct300868x,
C7CP04913G,doi:10.1063/1.4971853,C6CP08761B,doi:10.1002/wcms.30,PhysRevLett.108.236402}. 
In spite of its successes, 
describing the order of stability and the energetics of low-lying water cluster ((H$_2$O)$_n$) isomer (e.g. 
for the hexamer ($n=6$) 
and pentamer ($n=5$)), 
and different ice phases are great challenges for semilocal DFT~\cite{doi:10.1063/1.4944633}. 
There have been several research works devoted, mainly on the accuracy of a 
particular theoretical method for liquid water and ice 
phases \cite{doi:10.1063/1.2790009,doi:10.1063/1.4893377,Chen10846,
doi:10.1063/1.5023611,doi:10.1063/1.2790009,doi:10.1063/1.4893377,doi:10.1063/1.5023611,doi:10.1063/1.1630560,
doi:10.1063/1.1782074,doi:10.1063/1.5023611}. Experimentally, also, a lot of attention has been paid on the binding 
nature and conformal ranking of different low-lying water isomer and ice phases \cite{doi:10.1063/1.4944633,
doi:10.1021/jp950696w,doi:10.1063/1.480775,doi:10.1063/1.481613,doi:10.1063/1.4893377}. 

Physically, the interactive nature of the water is hydrogen (H) bonded. In the H-bond, a non-negligible electronic 
charge overlap happens between the bonding constituents, whose 
interaction nature is neither fully density overlap (like covalent bonds) nor weak van-der-Waals (vdW) forces. In 
H-bonding both the charge transfer (CT) and 
dipole interactions (DI) effects are observed~\cite{doi:10.1021/ct200329e}. The H-bond holds the water in its liquid, 
gas or crystal ice form, and the H-bonding potential energy surfaces can be modelled using several methods based on 
the properties of the electronic density, such as 
the theory of atoms in molecules \cite{bader1981quantum,becke2007quantum,grabowski2001ab}, and the noncovalent 
interaction index \cite{contreras2011analysis,fabiano2014wave}.

Based on several 
experimental and theoretical confirmations, there exist five to eight low-lying energy isomer of the water hexamer in 
nature~\cite{TSAI1993181,LAASONEN1993208,doi:10.1021/ct600366k}. Moreover, isomer for the trimer, tetramer, pentamer and higher order 
cluster are also found~\cite{doi:10.1021/jp055127v,doi:10.1063/1.1813431,doi:10.1021/jp0570770}. Besides, several phases of ice have been also found
~\cite{PhysRevLett.69.462,PhysRevLett.105.195701,PhysRevLett.107.185701}. 
However, the correct prediction of the 
ordering, stability and most preferred structure of 
the water isomer and ice phases, is a difficult problem for theoretical methods 
\cite{doi:10.1063/1.4944633,doi:10.1063/1.3012573,doi:10.1021/jp077376k}. In this regard, several 
ab-initio studies have been performed on the level of accurate wavefunction (M{\o}ller-Plesset perturbation theory 
(MP2) and quasi-perturbative connected triple 
excitations 
(CCSD(T)))~\cite{doi:10.1063/1.4944633,doi:10.1021/jp8105919,doi:10.1063/1.4807330,doi:10.1021/jp104822e}, diffusion 
Monte Carlo (DMC)~\cite{doi:10.1063/1.3012573} 
and density functional \cite{doi:10.1063/1.4944633,doi:10.1063/1.3012573,Chen10846,Sun2016}. 
The stable isomer of the water hexamer~\cite{doi:10.1021/jp8105919} are the prism, cage, book1, book2, bag, cyclic-chair,
cyclic-boat2, and cyclic-boat2, and those are utilized for benchmark performance of different
wavefunction and density functional  based
approaches \cite{doi:10.1063/1.4944633,doi:10.1063/1.3012573,doi:10.1021/jp8105919,doi:10.1063/1.4807330,
doi:10.1021/jp104822e}. Though the CCSD(T), MP2 and DMC predict correctly the order of stability  of
these water isomer, their computational costs are expensive.
Hence, the most preferred method is the density functional semilocal approach.

Regarding the performance of semilocal DFT for water, different xc functionals behave differently to describe 
the H-bond, which is relatively
weaker than the covalent interaction but stronger than the pure dispersion bond, being comparable in magnitude with 
the CT and DI interactions. 
Moreover, in a system like water, there is
a competition between the covalent, H-bond and dispersion bonds~\cite{doi:10.1063/1.4944633} and a good
semilocal or vdW corrected functional must be able to
describe all these bonds in order to perform quite conventionally for water.

In several previous studies, the bare semilocal xc functionals and long-range corrected hybrids, together 
with their dispersion corrected versions have been tested
for the water hexamer~\cite{doi:10.1063/1.3012573,
Chen10846,doi:10.1021/jp077376k,doi:10.1063/1.4807330}.
Considering the recent progress on semilocal DFT, it was shown that the 
Strongly Constrained and Appropriately 
Normed Semilocal Density Functional (SCAN) \cite{PhysRevLett.115.036402} functional performs quite well for the 
different phases of water, isomer and ice, being also very accurate for the polarizability, dipole moments and 
diffusion coefficient \cite{Chen10846}.
It can accurately describe the short and intermediate range of non-covalent interactions, 
being capable to capture the H-bonding nature of the water. 
However, SCAN overestimates the total energies of the 
water hexamer isomer and cluster, comparing with 
the MP2 
\cite{doi:10.1021/jp8105919,doi:10.1063/1.4807330,doi:10.1021/jp104822e,doi:10.1021/jp077376k,
doi:10.1063/1.3012573}, and CCSD(T) results \cite{doi:10.1021/ct600366k}. 
Note that the MP2 and CCSD(T) within complete basis set limit (CBS) methods 
are considered as the state-of-art for the reference energies of water isomer, and they are using for
benchmarking xc functionals \cite{doi:10.1021/jp8105919,doi:10.1063/1.4807330,
doi:10.1021/jp104822e}. 

In this paper, we investigate the behavior of the recently proposed meta-generalized gradient approximations 
(meta-GGAs) for the 
relative energies of the different low-lying water isomer
and cluster, conformal ranking, vibrational frequencies, and properties of ice phases. Keeping in mind that there 
are several previous studies~\cite{doi:10.1063/1.4944633}, we consider only selected 
semilocal xc functionals in this work. We restrict ourself only on the level of semilocal approximations 
because of their simplicity and quick output for studying 
large water cluster. We are mainly focusing on the performance of the recently proposed Tao-Mo 
(TM)~\cite{PhysRevLett.117.073001} meta-GGA and 
its revised form (revTM)~\cite{doi:10.1021/acs.jpca.9b02921}. For comparison, the PBE and SCAN results are also 
included. The present paper is arranged as follows: In the 
next section, we describe the methods and showing a brief outline of the TM and revTM functionals. Following this, we 
discuss the obtained results and comparison of different functionals. 
We summarize and conclude our results lastly.   
\begin{figure}
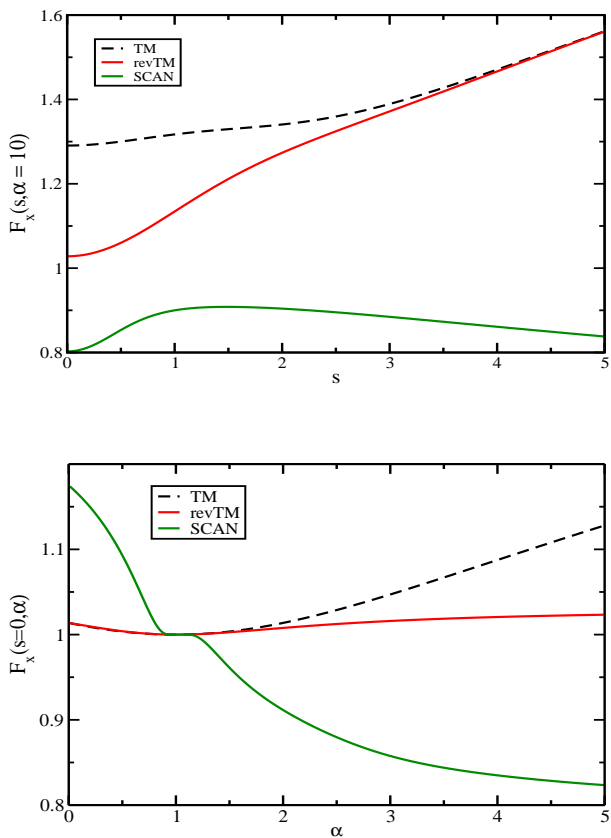

\begin{center}
\includegraphics[width=8.0cm,height=5.0cm]{enhence-plot.eps}
\\
\vspace{1 cm}
\includegraphics[width=8.0cm,height=5.0cm]{enhance-alpha.eps}
\end{center}
\caption{Exchange enhancement factor $F_x(s,\alpha)$ of TM, revTM, and SCAN meta-GGA functionals as a function of 
the reduced gradient $s$ (upper panel) and of the Pauli kinetic energy enhancement factor $\alpha$ (lower panel).}
\label{exen}
\end{figure}
\section{Theory}
Our starting point is the construction of the TM~\cite{PhysRevLett.117.073001} and 
revTM~\cite{doi:10.1021/acs.jpca.9b02921} functionals. In DFT, the xc energy functional is defined as,
\begin{eqnarray}
    E_{xc}&=&\int~d\bf{r}~\rho(\bf{r})~\epsilon_{xc}(\bf{r})~,
\end{eqnarray}
where $\rho$ is the electron density and $\epsilon_{xc}(\bf{r})$ is the xc energy per particle. The meta-GGA  
exchange functional is,
\begin{equation}
    E_x=\int~d{\bf{r}}~\rho({\bf{r}})~\epsilon^{unif}_{x}({\bf{r}})F_x(\rho,\nabla\rho,\tau^{KS})~,
\end{equation}
where $\tau^{KS}$ is the KS kinetic energy density, $F_x$ is the exchange enhancement factor, and 
$\epsilon^{unif}_{x}$ is the exchange energy per particle of the uniform electron gas. In 
Ref.~\cite{doi:10.1021/acs.jpca.9b02921}, the construction of 
the revTM is discussed in detail along with the TM functional. 
Here, we only focus on the key differences of the two functionals. The TM exchange enhancement factor 
\cite{PhysRevLett.117.073001} is given by
\begin{eqnarray}\label{final}
F_x^{TM} = w F_x^{DME} + (1-w) F_x^{sc}~,
\end{eqnarray}
where $w$ is a weighting factor, $F_x^{DME}$ is the density matrix expansion (DME) based semilocal exchange 
enhancement factor and 
$F_x^{sc}$ is the enhancement factor of the fourth-order gradient expansion (GE4) exchange energy. The DME part is 
accurate for the 
localized systems such as atoms and molecules, whereas the GE4 is accurate for slowly varying solid-state systems. 
See Refs. \cite{PhysRevLett.117.073001,doi:10.1021/acs.jpca.9b02921} for details.

In the revTM functional, the DME part remains unaltered, whereas the slightly different form of the $F_x^{sc}$ is 
proposed. The $F_x^{sc}$ of TM functional is given by   
\begin{equation}
\begin{split}
 F_x^{sc} = \{1+10[(10/81+50p/729)p+
146{\tilde q}^2/2025\\
-(73{\tilde q}/405)[3\tau_W/(5\tau)](1-\tau_W/\tau)]\}^{1/10}~,
\end{split}
\end{equation}
where $p=s^2$ with $s=|\nabla \rho|/[2(3\pi^2)^{1/3}\rho^{4/3}]$ being the reduced gradient, and $\tilde{q} = 
(9/20)(\alpha - 1) +  2p/3$ is finite near the nucleus (where $\alpha\approx 0$ and $s\approx 0.4$) and otherwise 
mimics the reduced Laplacian of the density (a.i. $q=|\nabla^2 
\rho|/[4(3\pi^2)^{2/3}\rho^{5/3}]$) outside of the nuclear region. 
Here $\alpha=\frac{\tau-\tau_W}{\tau^{unif}}$ is the Pauli kinetic energy enhancement factor used as a meta-GGA 
ingredient, being also an iso-orbital indicator, $\tau_W=\frac{|\nabla\rho|^2}{8\rho}$ is the von 
Weizs\"{a}cker kinetic energy density, and $\tau^{unif}$ is the kinetic energy density of the uniform electron gas. 
The difference of the TM and revTM in the exchange part comes from the 
construction of the $\tilde{q} $. Instead of the above defined $\tilde{q} $ as used in 
the TM functional, in revTM a slightly modified  $\tilde{q}_b = \frac{9(\alpha-1)}{20[1+b\alpha 
(\alpha-1)]^{1/2}}+\frac{2p}{3}$ is considered with $b=0.40$ \cite{doi:10.1063/1.1665298}. Due to this modification, 
the revTM functional behaves differently in the region $\alpha>>1$ (recognized as the overlapping closed 
shells~\cite{PhysRevLett.111.106401}) and $s\approx 0$. The $\alpha>>1$ region, where $F_x^{revTM}\le F_x^{TM}$ by 
construction, is found in the bonds of the H-bond, CT, and DI complexes. 

In Fig.~\ref{exen} we show a comparison of the TM, revTM, and SCAN meta-GGA enhancement factors. In the upper panel we 
plot $F_x(s,\alpha=10)$ versus $s$. The difference between TM and revTM is decreasing when $s$ increases, being  
maximum at $s=0$, and vanishing at $s\approx 3.5$. However, both TM and revTM exchange enhancement factors increase 
monotonically, enhancing over the Local Density Approximation (LDA) (note that $F_x^{LDA}=1$). Oppositely, SCAN shows 
important de-enhancement ($F_x^{SCAN}\le 1$), with a hill-like structure. Large values of $\alpha$, 
such as $\alpha=10$, are present in non-covalent bonds and in the tail of the density due to the degenerate 
valence shells.

In the lower panel of Fig.~\ref{exen}, we plot $F_x(s=0,\alpha)$ versus $\alpha$. 
The revTM remains almost steady for the whole range of the $\alpha$, whereas the TM
monotonically increases from $\alpha\ge 1$, and SCAN decreases. At $\alpha=0$, TM and revTM are close to LDA, while 
SCAN 
approaches its maximum value of 1.174 \cite{PhysRevLett.115.036402}. At $\alpha=1$, all functionals recover, by 
construction, the LDA behavior.   
Note that always $s=0$ in the middle 
of any bond, while $\alpha\approx 1$ for strong bonds (e.g. covalent bonds), $\alpha\ge 1$ for weak bonds (e.g. 
non-covalent bonds), and $\alpha\approx 0$ in the bonds where the valence atomic shells are non-degenerate (e.g. 
H$_2$, Be$_2$). 

The TM correlation functional is just a modified TPSS correlation with an improved performance in the low-density 
limit \cite{PhysRevLett.117.073001}. The $\beta=0.066725$ parameter
used in the TM functional is same as TPSS and PBE, representing the second-order coefficient of the 
correlation energy gradient expansion in the high-density limit. But, in the 
revTM $\beta$ is changed to
$\beta^{revTPSS}(r_s)=0.066725(1+0.1r_s)/(1+0.1778r_s)$, which is the correct, density-dependent second-order 
coefficient. Here $r_s$ is the bulk parameter (also called Wigner-Seitz radius). Apart from this modification,
the revTM keeps all the good properties of the TM correlation energy functional. Note that the change in the
correlation part bit improves the atomization energies of
molecules~\cite{doi:10.1021/acs.jpca.9b02921}.

\begingroup
\squeezetable
\begin{table*}[t]
\caption{\label{tab5ll} The reference values of the interaction energies, and the deviations (result - reference) 
obtained from several xc functionals, for the HB6 
\cite{Peverati20120476,doi:10.1021/ct6002719}, 
DI6 \cite{Peverati20120476,doi:10.1021/ct6002719} and CT7 \cite{Peverati20120476,doi:10.1021/ct6002719} representative 
test sets. The 6-311++G(3df,3pd) basis set within NWCHEM code is used 
\cite{doi:10.1021/acs.jpca.9b02921}. All values 
are in kcal/mol. We also report the mean error (ME) and mean absolute error (MAE) for each test, and the overall total 
ME (TME), and total MAE (TMAE). Best (T)MAE is in bold style.}
{
\begin{ruledtabular}
\begin{tabular}{cccccccccccccccccccccccccccccccccc}
Complex	&	Ref.&PBE &	SCAN	&	TM	&	revTM	\\
\hline
\multicolumn{6}{c}{HB6 (kcal/mol)} \\[0.2 cm]
NH$_3$-NH$_3$	&	3.1	&	0.0	&	0.3	&	0.0	&	-0.2	\\
HF-HF	&	4.6	&	0.3	&	1.1	&	0.4	&	0.1	\\
H$_2$O-H$_2$O	&	5.0	&	0.2	&	0.7	&	0.1	&	0.0	\\
NH$_3$-H$_2$O	&	6.4	&	0.5	&	0.6	&	0.1	&	0.1	\\
HCONH$_2$-HCONH$_2$	&	14.9	&	-0.6	&	1.1	&	-0.2	&	-0.7	\\
HCOOH-HCOOH	&	16.1	&	0.3	&	1.8	&	0.4	&	0.0	\\
	&		&		&		&		&		\\
ME	&		&	0.1	&	0.9	&	0.1	&	-0.1	\\
MAE	&		&	0.3	&	0.9	&	\textbf{0.2}	&	\textbf{0.2}	\\
\\[0.2 cm]
\multicolumn{6}{c}{DI6 (kcal/mol)} \\[0.2 cm]
H$_2$S-H$_2$S	&	1.7	&	0.0	&	-1.7	&	0.0	&	-0.2	\\
HCl-HCl	&	2	&	-0.1	&	0.6	&	0.1	&	-0.1	\\
HCl-H$_2$S	&	3.4	&	-0.7	&	0.2	&	0.5	&	0.4	\\
CH$_3$Cl-HCl	&	3.5	&	0.1	&	1.7	&	0.5	&	-0.1	\\
HCN-CH$_3$SH	&	3.6	&	0.1	&	0.8	&	0.2	&	-0.2	\\
CH$_3$SH-HCl	&	4.2	&	-1.4	&	2.0	&	1.7	&	1.2	\\
	&		&		&		&		&		\\
ME	&		&	-0.3	&	0.6	&	0.5	&	0.2	\\
MAE	&		&	\textbf{0.4}	&	1.2	&	0.5	&	\textbf{0.4}	\\
\\[0.2 cm]
\multicolumn{6}{c}{CT7 (kcal/mol)} \\[0.2 cm]
C$_2$H$_4$-F$_2$	&	1.06	&	2.14	&	1.84	&	2.14	&	1.84	\\
NH$_3$-F$_2$	&	1.81	&	3.49	&	2.59	&	3.19	&	2.89	\\
C$_2$H$_2$-ClF	&	3.81	&	2.29	&	3.49	&	2.99	&	2.59	\\
HCN-ClF	&	4.86	&	0.94	&	0.64	&	1.24	&	0.94	\\
NH$_3$-Cl$_2$	&	4.88	&	3.02	&	2.52	&	3.02	&	2.82	\\
H$_2$O-ClF	&	5.36	&	1.84	&	2.34	&	2.44	&	2.14	\\
NH$_3$-ClF	&	10.62	&	6.28	&	6.48	&	5.78	&	5.88	\\
	&		&		&		&		&		\\
ME	&		&	2.86	&	2.84	&	2.97	&	2.73	\\
MAE	&		&	2.86	&	2.84	&	2.97	&	\textbf{2.73}	\\
[0.2 cm]
\hline
TME	&		&	0.88	&	1.46	&	1.20	&	0.93	\\
TMAE	&		&	1.19	&	1.65	&	1.22	&	\textbf{1.09}	\\
\end{tabular}
\end{ruledtabular}
}
\end{table*}
\endgroup
In order to better understand and quantify the differences between the considered functionals, we report in Table 
\ref{tab5ll} a comparison between PBE, SCAN, TM and revTM for the H-bond (HB6 test set \cite{doi:10.1021/ct6002719}),  
dipole-dipole interaction (DI6 test set \cite{doi:10.1021/ct6002719}), and charge 
transfer interaction (CT7 test set \cite{doi:10.1021/ct6002719}). All these tests (HB6, DI6, and CT7) are 
representative, well-known and widely used. For all the systems, with the exception of NH$_3$-ClF 
charge-transfer complex, the revTM gives smaller deviations than TM, mainly due to the less exchange enhancement shown 
in Fig. \ref{exen}. Moreover, for all the tests, revTM improves over TM, showing the best performance, and implicitly 
the best overall error (TMAE=1.09 kcal/mol).    

\begingroup
\squeezetable
\begin{table*}[t]
\caption{\label{tab1}Binding energies per monomer (in meV/H$_2$O) of four water hexamer obtained from different 
functionals. 
The reference MP2, CCSD(T) and DMC reference values are taken from 
Refs.~\cite{doi:10.1063/1.3012573}. The MP2 
optimized geometries~\cite{doi:10.1063/1.3012573} are considered for single-point energy calculations. The 
aug-cc-pVTZ basis set is used. The most stable structure is in bold style.}
{
\begin{ruledtabular}
\begin{tabular}{cccccccccccccccccccccccccccccccccc}
Methods	&	Prism	&	Cage	&	Book	&	Cycle	\\
\hline
MP2&\textbf{332.3}&331.9&330.2&324.1\\
DMC&\textbf{331.9}&329.5&327.8&320.1\\
CCSD(T)&\textbf{347.6}&345.5&338.9&332.5\\[0.2 cm]
PBE&336.1&338.1&\textbf{344.0}&341.7\\
E$_{PBE}$-E$_{MP2}$&3.8&6.2&13.8&17.6\\
E$_{PBE}$-E$_{DMC}$&4.2&8.6&16.1&21.6\\
E$_{PBE}$-E$_{CCSD(T)}$&-11.5&-7.4&5.1&9.2\\[0.2 cm]
SCAN&\textbf{377.4}&375.5&369.7&359.3\\
E$_{SCAN}$-E$_{MP2}$&45.0&43.6&39.5&35.2\\
E$_{SCAN}$-E$_{DMC}$&45.5&46.0&41.9&39.2\\
E$_{SCAN}$-E$_{CCSD(T)}$&29.8&30.0&30.8&26.8\\[0.2 cm]

TM&\textbf{354.7}&342.6&330.6&316.8\\
E$_{TM}$-E$_{MP2}$&22.4&10.7&0.34&-7.3\\
E$_{TM}$-E$_{DMC}$&22.8&13.1&2.8&-3.3\\
E$_{TM}$-E$_{CCSD(T)}$&7.1&-2.9&-8.3&-15.7\\[0.2 cm]

revTM&\textbf{333.1}&331.4&328.0&320.0\\
E$_{revTM}$-E$_{MP2}$&0.8&-0.5&-2.2&-4.1\\
E$_{revTM}$-E$_{DMC}$&1.2&1.9&0.2&-0.1\\
E$_{revTM}$-E$_{CCSD(T)}$&-14.5&-14.1&-10.9&-12.5\\

\end{tabular}
\end{ruledtabular}
}
\end{table*}
\endgroup
\begin{figure}
\begin{center}
\includegraphics[width=9.0cm,height=6.0cm]{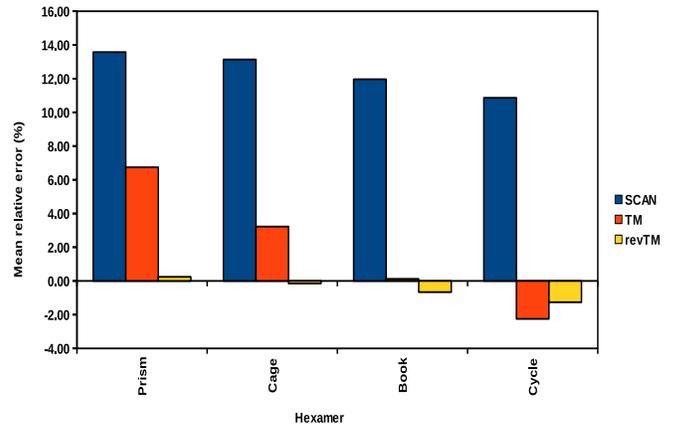}
\end{center}
\caption{Shown is the percentage deviation in binding energy per monomer,
for different meta-GGA functionals compared to the MP2 reference, of the four
water hexamer shown in Table~\ref{tab1}.}
\label{fig1}
\end{figure}
\begingroup
\squeezetable
\begin{table*}
\caption{\label{tab2}Relative binding energies (with respect to the prism structure) for different low-lying 
energy isomer of the water hexamer. The reference values are from high-level 
wavefunction calculations (CCSD(T)/CBS of Ref.~\cite{doi:10.1021/jp8105919}). The MP2/aug-cc-pVTZ level geometries 
\cite{doi:10.1021/jp104822e}, and the aug-cc-pVTZ basis set are used.
All results are in kcal/mol. Best calculated values are in bold style.}
{
\begin{ruledtabular}
\begin{tabular}{cccccccccccccccccccccccccccccccccc}
Methods	&	Cage	&	Book1	&	Book2	&	Bag	&	Chair	&	Boat1	&	
Boat2&ME&MAE	\\
\hline
Ref.	&	0.25	&	0.72	&	1.02	&	1.62	&	1.80	&	2.79	&	
2.85&$-$&$-$	\\[0.2 cm]
PBE	&	-0.27	&	-1.07	&	-0.81	&	-0.06	&	-0.76	&	0.29	&	
0.46&-1.90&1.90	\\
SCAN	&	0.25	&	1.05	&	1.34	&	1.91	&	2.50	&	3.56	&	
3.69&0.46&0.46	\\
TM	&	0.66	&	2.32	&	2.58	&	3.08	&	4.22	&	5.17	&	
5.28&1.75&1.75	\\
revTM	&	\textbf{0.24}	&	\textbf{0.71}	&	\textbf{0.98}	&	\textbf{1.68}	&	\textbf{1.82}	&	\textbf{2.81}	&	
\textbf{2.94}&0.02&\textbf{0.03}	\\
\end{tabular}
\end{ruledtabular}
}
\end{table*}
\endgroup
\begingroup
\squeezetable
\begin{table*}
\caption{\label{jpca1} Comparison of relative binding energies with respect to the most stable structure (a.i. 4-S$_4$ 
for tetramer and 5-CYC for pentamer) of 
water tetramer and pentamer low-lying energy isomer ~\cite{doi:10.1021/jp2069489}, using different methods. 
The CCSD(T)/CBS reference values are taken from~\cite{doi:10.1021/jp2069489}. The RI-MP2/aVDZ optimized 
coordinates \cite{doi:10.1021/jp2069489} and the aug-cc-pVTZ basis set are used. All 
results are in kcal/mol. Best calculated values are in bold style.}
{
\begin{ruledtabular}
\begin{tabular}{c||cc||ccccccccccccccccccccccccccccccc}
	&	4-Ci&4-Py&5-FR-B	&	5-CA-C	&	5-CA-A	&	5-CA-B	&	5-FR-A	&	5-FR-C	\\
	\hline
Ref.	&	0.85	&	3.55&	1.14	&	1.32	&	1.47	&	2.19	&	2.88	&	
3.57	\\[0.2 cm]
PBE	&0.89&5.16&	2.42	&	3.32	&	3.65	&	4.07	&	4.62	&	\textbf{3.95}	\\
SCAN	&0.95&3.79&	1.15	&	1.04	&	1.29	&	1.97	&	3.53	&	2.75	\\
TM	&0.89&2.47&	0.15	&	-0.58	&	-0.38	&	0.30	&	2.36	&	1.62	\\
revTM	&\textbf{0.88}&\textbf{3.66}&	\textbf{1.13}	&	\textbf{1.21}	&	\textbf{1.46}	&	\textbf{1.98}	&	\textbf{3.28}	&	2.59	\\
\end{tabular}
\end{ruledtabular}
}
\end{table*}
\endgroup
\begin{figure}
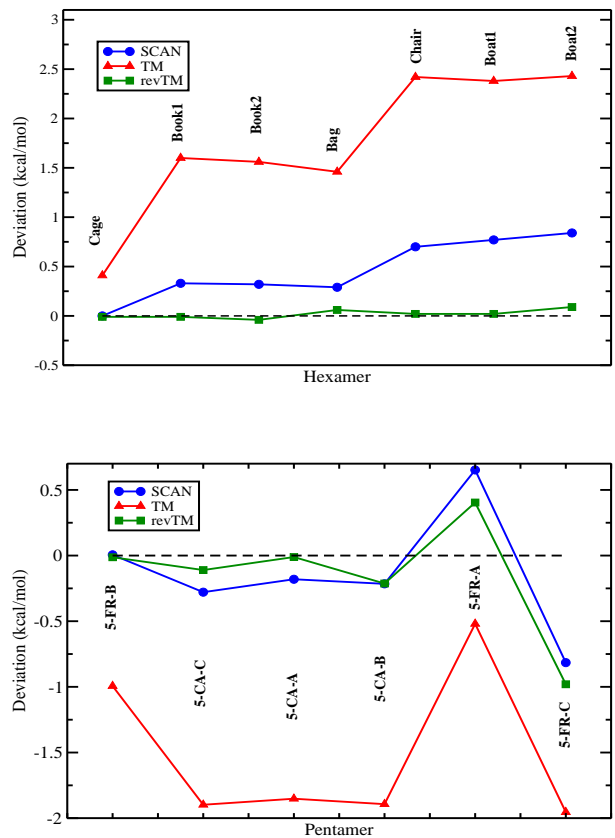

\begin{center}
\includegraphics[width=8.0cm,height=5.0cm]{water-hex-2.eps}\\
\vspace{1.0 cm}
\includegraphics[width=8.0cm,height=5.0cm]{pentemer.eps}
\end{center}
\caption{Deviation (result - reference) of the different functionals for the water hexamer of Table \ref{tab2} (upper 
panel) and water pentamer of Table \ref{jpca1} (lower 
panel). Only meta-GGA functionals are considered here.}
\label{hex-plot}
\end{figure}
\begingroup
\squeezetable
\begin{table*}[t]
\caption{\label{tab3}Comparison of different functionals for four different stable water cluster ((H$_2$O)$_n$, 
with $n=2$ to 5). 
The MP2/CBS reference dissociation energies per monomer (in meV/H$_2$O) are taken from~\cite{doi:10.1063/1.2790009}. 
The single point energy calculations are done with MP2/aug-cc-pVTZ level geometries~\cite{doi:10.1063/1.2790009}. 
In parentheses we report the percentage deviation from the dissociation energy of the dimer. The aug-cc-pVTZ 
basis set is used. Best calculated values are in bold style.}
{
\begin{ruledtabular}
\begin{tabular}{cccccccccccccccccccccccccccccccccc}
Methods	&	Dimer	&	Trimer	&	Tetramer	&	Pentamer&ME&MAE	\\
\hline
MP2	& 215.8&228.5 (5.9)&299.9 (38.9)&314.4 (45.7)&$-$&$-$	\\[0.2 cm]
PBE	&219.2 &231.1 (5.4)&314.3 (43.4)&332.7 (51.8)& 9.7&9.7\\
SCAN	&235.1  &256.6 (9.1)&335.9 (42.8)&350.8 (49.2)&30.0&30.0\\
TM	&\textbf{217.2} &241.1 (11.0)&306.5 (41.1)&317.0 (46.0)&5.8&5.8\\
revTM	&209.4 &\textbf{228.5} (9.1)&\textbf{299.0} (42.8)&\textbf{312.2} (49.1)&-2.3&2.4\\
\end{tabular}
\end{ruledtabular}
}
\end{table*}
\endgroup
\begingroup
\squeezetable
\begin{table*}[t]
\caption{\label{tab5}The reference values and the deviations (result - reference) obtained from xc functionals, for 
the WATER27 test set~\cite{C7CP04913G}. The CCSD(T)-F12/CBS reference values are taken 
from Ref. \cite{manna2017conventional}. 
The def2-QZVPD basis set is used. All values are in kcal/mol. Best MAE are in bold style.}
{
\begin{ruledtabular}
\begin{tabular}{cccccccccccccccccccccccccccccccccc}
Complex	&	Ref.&PBE &	SCAN	&	TM	&	revTM	\\
\hline
\multicolumn{6}{c}{neutral (kcal/mol)} \\[0.2 cm]
(H$_2$O)$_2$	&4.974&	0.082	&	0.420	&	0.010	&	-0.167	\\
(H$_2$O)$_3$ cyclic	&15.708&	0.262	&	1.903	&	0.850	&	0.008	\\
(H$_2$O)$_4$ cyclic	&27.353&	1.519	&	3.333	&	0.720	&	0.041	\\
(H$_2$O)$_5$ cyclic	&35.879&	2.299	&	4.211	&	0.459	&	-0.117	\\
(H$_2$O)$_6$ prism	&45.988&	0.096	&	5.351	&	2.351	&	-0.531	\\
(H$_2$O)$_6$ cage	&45.733&	0.697	&	5.412	&	2.029	&	-0.432	\\
(H$_2$O)$_6$ book	&45.292&	2.016	&	5.269	&	1.052	&	-0.291	\\
(H$_2$O)$_6$ cyclic	&44.296&	2.765	&	4.990	&	0.297	&	-0.306	\\
(H$_2$O)$_8$ cube (D$_{2d}$)	&72.490
&	1.492	&	8.215	&	2.427	&	-0.886	\\
(H$_2$O)$_8$ cube (S$_4$)	&72.454&	1.515	&	8.262	&	2.426	&	-0.875	\\
(H$_2$O)$_{20}$ dodecahedron	&197.804
&	9.163	&	21.009	&	1.101	&	-3.450	\\
(H$_2$O$)_{20}$ fused cubes	&207.534
&	-2.506	&	20.763	&	6.356	&	-5.775	\\
(H$_2$O$)_{20}$ face-sharing	&207.763
&	-0.464	&	20.259	&	4.029	&	-6.331	\\
(H$_2$O$)_{20}$ edge-sharing	&209.081&	1.415	&	20.687	&	3.835	&	-5.417	\\[0.2 cm]
ME&&1.454	&	9.292	&	1.996		&-1.752\\
MAE&&1.878&		9.292&		1.996&		\textbf{1.759}\\
\\[0.2 cm]
\multicolumn{6}{c}{protonated (kcal/mol)} \\[0.2 cm]
(H$_3$O)$^+$(H$_2$O)	&33.738&	3.169	&	3.040	&	0.404	&	0.525	\\
(H$_3$O)$^+$(H$_2$O)$_2$	&57.114&	3.262	&	4.037	&	0.171	&	0.284	\\
(H$_3$O)$^+$(H$_2$O)$_3$	&76.755&	2.924	&	4.142	&	-0.105	&	-0.110	\\
(H$_3$O)$^+$(H$_2$O)$_6$ (3D)	&117.683
&	3.142	&	7.246	&	0.905	&	-0.873	\\
(H$_3$O)$^+$(H$_2$O)$_6$ (2D)	&114.819
&	4.312	&	6.580	&	-0.367	&	-1.052	\\[0.2 cm]
ME&&3.362&		5.009&		0.201&		-0.245\\
MAE&&3.362&		5.009&		\textbf{0.390}&		0.569
\\[0.2 cm]
\multicolumn{6}{c}{deprotonated (kcal/mol)} \\[0.2 cm]
OH$^-$(H$_2$O)	&26.687&	2.165	&	2.797	&	0.187	&	0.137	\\
OH$^-$(H$_2$O)$_2$	&48.688&	1.983	&	3.785	&	0.005	&	-0.128	\\
OH$^-$(H$_2$O)$_3$	&67.525&	1.837	&	4.470	&	-0.068	&	-0.416	\\
OH$^-$(H$_2$O)$_4$ (C$_4$)	&84.351&	-0.426	&	5.035	&	0.799	&	-1.146	\\
OH$^-$(H$_2$O)$_4$ (C$s$)	&85.015&	0.458	&	6.243	&	1.457	&	-0.682	\\
OH$^-$(H$_2$O)$_5$	&100.782&	-0.995	&	7.138	&	2.663	&	-0.705	\\
OH$^-$(H$_2$O$)_6$	&115.672&	-0.526	&	8.485	&	2.719	&	-0.978	\\[0.2 cm]
ME&&0.643&		5.422&		1.109&		-0.560\\
MAE&&1.199&		5.422&		1.128&		\textbf{0.599}\\[0.2 cm]
\multicolumn{6}{c}{mixed (kcal/mol)} \\[0.2 cm]
(H$_2$O)$_8$ cube (S$_4$)$-$
H$_3$O$^+$(H$_2$O)$_6$OH$^-$	&29.605&	-12.003	&	-8.060	&	-0.963	&	-3.724	\\[0.2 cm]
ME&&-12.003	&	-8.060	&	-0.963	&	-3.724\\
MAE&&12.003	&	8.060	&	\textbf{0.963}	&	3.724\\[0.2 cm]
\hline
TME&&1.10 &6.85 &1.32 &-1.24\\
TMAE	&&2.35 &\underline{7.44} &1.44 &\textbf{1.31}	\\
\end{tabular}
\end{ruledtabular}
}
\end{table*}
\endgroup
\begingroup
\squeezetable
\begin{table}[t]
\caption{\label{tabs} Bond lengths (\AA), bond angle (degrees), dipole moment (Debye), and static 
polarizability (a.u.) of the water monomer. 
The experimental reference values are taken from Refs.\cite{doi:10.1021/ct200329e,gray1984theory}. The aug-cc-pVTZ 
basis set is used.}
{
\begin{ruledtabular}
\begin{tabular}{cccccccccccccccccccccccccccccccccc}
    &Expt.	&PBE &	SCAN	&	TM&revTM	\\
\hline
$R(O-H)$&0.957&0.970&0.961&0.969&0.969		\\
$\angle HOH$&104.5&104.2&104.4&103.9&104.0\\		
$\mu$&1.855&1.804&1.840&1.794&1.798\\
$\alpha$&9.92&10.33&9.81&10.29&10.28
\\
\end{tabular}
\end{ruledtabular}
}
\end{table}
\endgroup
\begingroup
\squeezetable
\begin{table}[t]
\caption{\label{tavviii} Mean absolute relative error (MARE) (in \%) of harmonic vibrational frequencies for water 
cluster and hexamer isomer, using different methods.
All values are in cm$^{-1}$ and calculated with an aug-cc-pVTZ basis set. Details of reference and individual values 
are given in Ref.~\cite{suppli}.}
{
\begin{ruledtabular}
\begin{tabular}{cccccccccccccccccccccccccccccccccc}
		 &PBE &	SCAN	&	TM&revTM	\\
\hline
H$_2$O &\underline{3.6}	&	\textbf{0.7}	&	2.8	&	2.8\\
(H$_2$O)$_2$&6.0	&	\underline{6.7}	&	4.1	&	\textbf{2.6} \\
(H$_2$O)$_3$&\underline{8.6}	&	8.5	&	\textbf{2.7}	&	4.9 \\
(H$_2$O)$_4$&\underline{9.7}	&	9.1	&	\textbf{3.0}	&	3.2 \\
(H$_2$O)$_5$&12.7	&	\underline{15.8}	&	\textbf{3.6}	&	5.1 \\
(H$_2$O)$_6$-book&\underline{10.7}	&	8.5	&	\textbf{3.3}	&	3.9 \\
(H$_2$O)$_6$-cage&\underline{8.8}	&	6.1	&	4.0	&	\textbf{3.5} \\
(H$_2$O)$_6$-prism&\underline{7.1}	&	6.3	&	3.8	&	\textbf{2.8} \\
(H$_2$O)$_6$-ring&\underline{11.8}	&	10.5	&	\textbf{3.0}	&	4.3 \\
\end{tabular}
\end{ruledtabular}
}
\end{table}
\endgroup
%
%
%
%
\begin{figure*}
\begin{center}
\includegraphics[width=17.0cm,height=7.0cm]{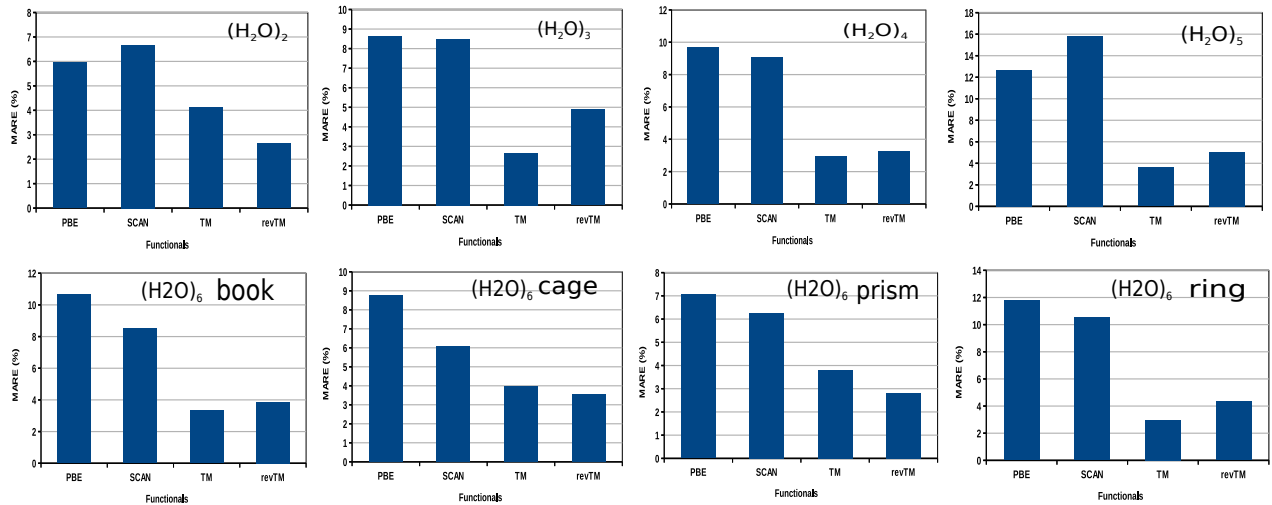}
\end{center}
\caption{MAREs of vibrational frequencies obtained from different xc functionals, for the water cluster and isomer
of Table \ref{tavviii}.}
\label{mare-plot}
\end{figure*}
%
\begin{figure}
\begin{center}
\includegraphics[width=6.0cm,height=5.0cm]{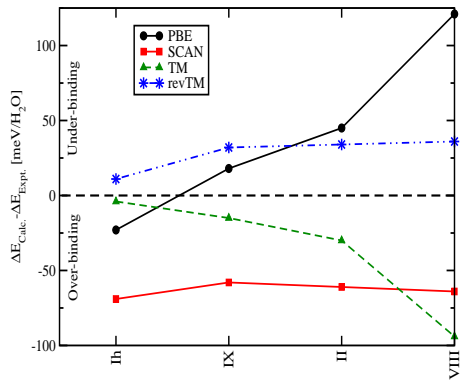}
\end{center}
\caption{Deviations in the formation energies of several ice phases, obtained from different xc functionals. (See 
Table \ref{ice1}.)}
\label{ice-1}
\end{figure}
%
%
\begin{figure}
\begin{center}
\includegraphics[width=6.0cm,height=5.0cm]{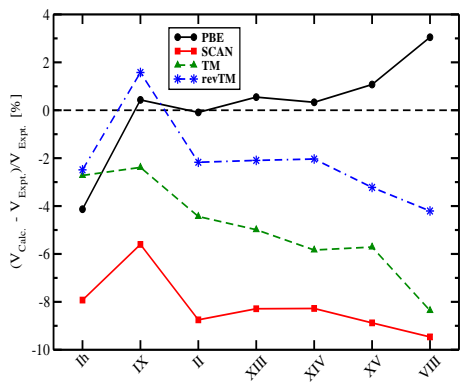}
\end{center}
\caption{Shown is the relative differences in the volume of ice as calculated using different methods. (See
Table \ref{ice2}.)}
\label{ice-2}
\end{figure}
\begingroup
\squeezetable
\begin{table*}
\caption{\label{ice1} Sublimation energies per monomer of different phases of ice, computed with various methods.
The relative energies of each structure with respect to the ice I$h$, are given in parentheses.
All energies are in meV/H$_2$O. Experimental and DMC values from Ref.~\cite{doi:10.1063/1.4824481}. 
}
{
\begin{ruledtabular}
\begin{tabular}{cccccccccccccccccccccccccccccccccc}
           & I$h$      & IX        & II        & XIII      & XIV       & XV         & VIII    \\
\hline
Expt.      & -610 (0)  & -606 (5)  & -609 (1)  &  --       &   --      &   --       & -577 (33) \\
DMC        & -605 (0)  & --        & -609 (-4) &  --       &   --      &   --       & -575 (30) \\[0.2 cm]

PBE        &-633 (0)   &-588 (45)  &-564 (69)  &-553 (80)  &-540 (93)  &-523 (110)  &-456 (177) \\
SCAN       &-679 (0)   &-664 (15)  &-670 (9)  &-660 (19)  &-658 (21)  &-652 (27)  &-641 (38) \\
TM         &-614 (0)   &-621 (-7)  &-639 (-25)  &-632 (-18)  &-635 (-21)  &-638 (-24)  &-671 (-57) \\
revTM      &-599 (0)   &-574 (25)  &-575 (24)  &-565 (34)  &-560 (39)  &-554 (45)  &-541 (58) \\
\end{tabular}
\end{ruledtabular}
}
\end{table*}
\endgroup
\begingroup
\squeezetable
\begin{table*}
\caption{\label{ice2} Calculated and experimental equilibrium volumes (\AA$^3$/H$_2$O) of different phases of ice. 
The zero-point-energy \cite{rasti2019importance} corrected experimental 
and DMC values are taken from Ref.~\cite{doi:10.1063/1.4824481}. 
}
{
\begin{ruledtabular}
\begin{tabular}{cccccccccccccccccccccccccccccccccc}
           & I$h$        & IX        & II        & XIII      & XIV       & XV        & VIII          \\
\hline
Expt.      & 32.05 & 25.63 & 24.97 & 23.91 & 23.12 & 22.53 & 20.09    \\
DMC        
           & 31.69     &           & 24.70     &           &           &           & 19.46         \\   [0.2 cm]

PBE        &30.73&	25.74&	24.95&	24.04&	23.20&	22.77&	20.70 \\
SCAN       &29.51	&24.19	&22.79	&21.93	&21.21	&20.53	&18.19 \\
TM       &31.18	&25.02	&23.86	&22.72	&21.77	&21.24	&18.41 \\
revTM       &31.25&	26.03&	24.43&	23.41&	22.65&	21.80&	19.25 \\
\end{tabular}
\end{ruledtabular}
}
\end{table*}
\endgroup
\section{Results and Discussions}

\subsection{Water isomer}
In Table \ref{tab1}, we present the binding energies per monomer of four low-lying isomer of the water hexamer 
(prism, cage, book and cycle), obtained from the PBE, SCAN, TM and revTM functionals, using the MP2 optimized 
geometries \cite{doi:10.1063/1.3012573}.
For the reference energy benchmarks we 
consider the MP2 \cite{doi:10.1021/ct600366k,doi:10.1063/1.3012573,doi:10.1021/jp077376k}, CCSD(T)/CBS 
\cite{doi:10.1021/jp8105919}, and DMC \cite{doi:10.1063/1.3012573} values. Note 
that, the MP2 is the standard procedure for benchmarking the water hexamer as it is used in different 
works \cite{doi:10.1063/1.3012573,doi:10.1021/jp077376k,doi:10.1063/1.1423941,doi:10.1063/1.1864892,
doi:10.1021/jp054958y,doi:10.1063/1.4807330,doi:10.1021/jp104822e,doi:10.1021/ct600366k}.
Considering earlier attempts, it was shown that GGA (PBE) and GGA with vdW correction (GGA+vdW) functionals predict 
wrong energy ordering of the water hexamer \cite{doi:10.1063/1.3012573}, whereas, 
SCAN meta-GGA gives the correct ordering \cite{Sun2016}. In Table \ref{tab1}, it is shown that along with SCAN, the 
TM and revTM functionals also predict the correct ordering and stability 
(prism~$>$~cage~$>$~book~$>$~cycle) of these water hexamer. 
Moreover, the obtained energies from revTM functional agree closely 
with that of the MP2, DMC, and CCSD(T) results. On the other hand, the TM functional is close to the CCSD(T) 
results, while SCAN functional 
shows the most overestimation tendency within the 
considered functionals. The remarkable performance of the revTM functional is shown in 
Fig.~\ref{fig1}, where we also plot the relative errors (in \%) 
of SCAN, TM and revTM meta-GGAs. 

Next we focus on the relative stability of different low-lying isomer of the water hexamer, with 
respect to the prism structure, which is the most stable.
Our results are summarized in Table \ref{tab2}. Note that the prediction of accurate relative energies 
is still a long-standing problem.
Very recently, it was shown that the SCAN meta-GGA 
performs efficiently and predicts the correct 
order of stability, but the relative binding energies of each hexamer with respect to the prism structure,  are 
overestimated \cite{Chen10846,Sun2016,doi:10.1063/1.5023611,
PhysRevB.99.205123,doi:10.1063/1.5006146}. In our present study, we also found a similar tendency for the SCAN 
functional. However, from Table \ref{tab2}, we observe that 
the revTM functional performs remarkably and getting the absolute binding energies in very good agreement with 
the CCSD(T)/CBS reference. The revTM functional obtains the 
smallest mean absolute error (MAE=0.03 kcal/mol), that is an order of magnitude better than SCAN (MAE=0.46 kcal/mol).
Whereas, the TM functional 
tends to overestimate the binding energies (MAE=1.75 kcal/mol), but still gives the correct ordering between the 
hexamer. In the upper panel of Fig \ref{hex-plot}, we plot the deviations (result - reference) of meta-GGA 
functionals, for all considered isomer, showing that the revTM functional is almost exact.

To stress on the remarkable performance of revTM functional, in Table \ref{jpca1}, we also consider the relative 
binding energies of the low-lying water isomer for the tetramer ($n=4$) and
pentamer ($n=5$), and compare those with the CCSD(T)/CBS level theory~\cite{doi:10.1021/jp2069489}. For the pentamer 
structures, the cycle (5-CYC) is the most 
stable structure. In this case also, the revTM functional performs remarkable well and reproduce the correct ordering, 
with exception of the 5-FR-C pentamer where all functionals wrongly predict it to be before the 5-FR-A 
isomer. 
The SCAN functional is the second-best performer,  
while the TM functional fails not only to show the correct order, but also gives 5-CA-C as the most stable structure, 
instead of 5-CYC. This is an important shortcoming of the TM meta-GGA which revTM functional is able to solve it.
Finally, in the lower panel of Fig.~\ref{hex-plot}, we also plot the deviations in the relative
binding energies of these water pentamer. From this plot one can also observe that the revTM functional is
performing best compared to the others.

In the simpler case of the tetramer, all functionals show the correct order between the 4-S$_4$, 4-Ci, and 4-Py 
isomer. However, the relative binding energies between the 4-Ci and 4-Py are: 2.7 kcal/mol for reference, 4.27 
kcal/mol for PBE, 2.84 kcal/mol for SCAN, 1.58 kcal/mol for TM, and 2.78 kcal/mol for revTM.  

\subsection{Water cluster}
Let us now consider the small water cluster (with most stable structure) to assess the quality of the different 
functionals. The results of the dimer, trimer, tetramer, and pentamer are presented 
in Table \ref{tab3}. These small water cluster have been used to test the performance of 
several density functionals~\cite{doi:10.1063/1.2790009}. However, 
We observe the remarkable accuracy of the revTM functional with the smallest overall error (MAE=2.4 meV/H$_2$O),
followed by TM (MAE=5.8 meV/H$_2$O), and PBE (MAE=9.7 meV/H$_2$O), while SCAN strongly overestimates the dissociation 
energies giving MAE=30.0 meV/H$_2$O. 
To incorporate the H bond strength, in Table \ref{tab3} we also report, for each functional, the percentage deviation 
in dissociantion energy per monomer, of the trimer, tetramer, and pentamer with respect to the dimer.
From these results we observe that all the functionals perform equivalently. However, the revTM functional is a bit 
closer to the SCAN functional. 

\subsection{Performance for the WATER27 subset}
The remarkable accuracy of the revTM functional is also proved from Table \ref{tab5}, where we consider the binding 
energies of the 27 neutral, positively, and negatively charged water cluster (WATER27) from the 
well known GMTKN55 test set~\cite{C7CP04913G}. This test set consists of 
10 neutral (H$_2$O)$_n$ structures (with $2\le n\le 8$), 4 isomer of (H$_2$O)$_{20}$,
5 protonated water cluster of the form H$_3$O$^+$(H$_2$O)$_n$ (with $1\le n \le 6$),
7 deprotonated cluster of the form OH$^-$(H$_2$O)$_n$ (with $1\le n \le 6$), and 
1 mixed hydroxonium-hydroxide zwitterion.  
Results of Table \ref{tab5} indicate the astonishing performances of both TM and revTM with overall errors smaller 
than 1.5 kcal/mol, followed by PBE (TMAE=2.35 kcal/mol).
In this case, the SCAN functional deviates most (with TMAE=7.4 kcal/mol), showing a quite modest performance 
especially for the isomer of (H$_2$O)$_{20}$. 

\subsection{Structural and vibrational properties}
In Table \ref{tabs} we report the bond length, bond angle, dipole moment and static polarizability
of the water monomer.
We observe that TM and revTM functionals perform similarly, being slightly better than PBE, but noticeably worse than 
SCAN meta-GGA. Both TM and revTM overestimate the bond length and static polarizability, and 
underestimate the bond angle and dipole moment 

To assess the functionals for the harmonic vibrational frequencies of the water, we consider several 
water cluster (H$_2$O)$_{n}$ ($n = 1-5$), and isomer ($n = 6$). 
The full results are provided in Ref. 
\cite{suppli}, and here we summarize them in Table \ref{tavviii} and Fig~\ref{mare-plot}.  
As reference, we considered the CCSD(T) results from Refs.  
\cite{doi:10.1021/jp2069489,doi:10.1021/ct500860v,doi:10.1021/acs.jctc.5b00225}.
For the water monomer, we obtain the same trend as in Table \ref{tabs}, with SCAN being remarkably accurate (MARE= 
0.7\%), while TM (MARE=2.8\%) and revTM (MARE=2.8\%) being slightly better than PBE (MARE=3.6\%), which gives the 
worse performance. On the other hand, for all the other systems, from dimer to hexamer, SCAN works in line with, 
and sometimes worse than PBE, giving large errors (6.1 \% $\le$ MARE $\le$ 15.8\%). 
These results are also in agreement with Tables \ref{tab1} and \ref{tab3}, where SCAN gives significant overestimation 
of the binding energies of the water cluster. 
Finally, we observe that TM and revTM perform accurately for the harmonic vibrational frequencies, being able to 
describe all the systems within MARE$\le$5.1 \%. 
They are very good especially for lower
frequencies $\nu<1000$~cm$^{-1}$, whereas SCAN is the slightly better functional for
higher frequencies ($\nu>1000$~cm$^{-1}$). Note that PBE functional overestimates the vibrational
frequencies for $\nu<1000$~cm$^{-1}$ and underestimates the same for $\nu>1000$~cm$^{-1}$. 
 
\subsection{Performance for ice}
The ab-initio studies of ice phases are also quite important. Several experimental 
structures suggested that $15$ ice phases exist~\cite{C1CP21712G,
PhysRevLett.103.105701,PhysRevLett.107.185701,doi:10.1063/1.4824481}. Most of them are constructed from the proton 
disordered phases, having complex structures. However, only seven ice phases,
the ice Ih and six other proton ordered phases (IX, II, XIII, XIV, XV, and VIII) 
are considered to perform benchmark calculations for the semilocal, non-local and vdW corrected functionals 
\cite{PhysRevLett.107.185701,doi:10.1063/1.4824481}. 
Among all these phases the ice Ih is the
most stable structure. In this subsection we will report the performance of the 
different density functional approaches for the sublimation or lattice energies of the ice and zero pressure volumes. 

The sublimation energy per H$_2$O of ice phases, represents the energy per monomer required to change the ice into gas 
phase, and is defined as,
\begin{equation}
 \Delta E =(E^{Ice}-N\times E^{H_2O})/N~,
\end{equation}
where $E^{Ice}$ is the ice energy, $E^{H_2O}$ is the energy of the isolated H$_2$O molecules, and $N$ is the number 
of water molecules.  
Considering earlier works, several semilocal functionals and their vdW corrected versions have been 
studied~\cite{PhysRevLett.107.185701,
doi:10.1063/1.4824481,PhysRevB.87.214101}. It has been shown that all vdW corrected methods improve considerably 
the relative energies 
for the ice phases~\cite{doi:10.1063/1.4824481}. The results of the PBE, SCAN, TM and revTM functionals for the 
absolute and relative energies are 
presented in Table~\ref{ice1}, where for the benchmark calculations we used the experimental results and the DMC data 
taken from Ref.~\cite{doi:10.1063/1.4824481}. 

Inspection of Table \ref{ice1} shows that the revTM performs best for the total sublimation energies (see also 
Fig.~\ref{ice-1}), gives the correct ordering of the phases, but overestimates the relative energies with respect to 
the ice Ih. On the other hand, SCAN significantly underestimates the sublimation energies (see also
Fig.~\ref{ice-1}), but is remarkably accurate for the relative energies. Here we observe the collapse of the TM 
meta-GGA, that does not provide the correct ordering of the ice phases, gives negative relative energies, and
wrongly predicts the ice VIII as the most stable. In this respect, the performance of revTM is remarkable, 
correcting all these failures of the TM functional.    

The equilibrium volume is another important quantity used to assess the performance of different theoretical methods. 
Table~\ref{ice2} shows the equilibrium volumes of the ice phases.
We observe that the PBE functional underestimates the ice Ih
volume with about 4\% (see also Fig. \ref{ice-2}) and overestimates the higher density ice VIII with about 
3\%, otherwise being very accurate (below 1\%). 
Note that the present results of PBE functionals are consistent with 
the ones of Refs. \cite{PhysRevLett.107.185701,doi:10.1063/1.4824481}. 

However, for the SCAN and TM 
meta-GGAs, the tendendy is somehow different. Both the SCAN and TM functionals underestimate the equilibrium volume, 
from lower (Ih phase) to higher (VIII phase) 
densities. Presumably, this underestimation of the equilibrium volume is because of the presence of a bit much short or 
intermediate range vdW interaction in the functional form of both
of them. We recall that the vdW correction shrinks the volume \cite{PhysRevB.98.214108}. The SCAN and TM functionals 
underestimate the volume of the ice Ih by about 8\% and 3\%, and the volume of the ice VIII by about 9\% and 8\%, 
respectively (see also Fig. \ref{ice-2}). 
However, the underestimation tendency in volume of the TM functional is greatly modified by revTM
due to the de-enhancenment of the short-range vdW interaction in the revTM. 
To encapsulate the behavior of the different functionals, we show in 
Fig.~\ref{ice-2}, the relative deviations of the functionals, with respect to the experimental volumes, for several 
ice phases. PBE and revTM are the most accurate. Note that revTM results have a smaller standard deviation than the PBE 
ones.


\subsection{Ice lattice mismatch problem}
\label{subd}
The impressive performance of the revTM for different ice phases can also endeavor from the popular ice lattice 
mismatch problem. The lattice mismatch $(f)$ of ice Ih and $\beta$-AgI is defined as,
\begin{equation}
f= \frac{2(b-a)}{b+a}\times 100 \%\ ~,
\end{equation}
where $a$ and $b$ are the lattice constants of ice Ih and $\beta$-AgI respectively. 

Practically, $f$ accounts for many applications having industrial uses. For example, it determines the rate of growth 
of ice on a $\beta$-AgI surface. However, computationally one have to get both the lattice constants of ice Ih and 
$\beta$-AgI accurately to obtain the accurate prediction of $f$. This is quite a difficult task since it involves 
the simultaneous determination of both the lattice constants of ice Ih and $\beta$-AgI having different electronic 
properties. Most precisely, this is quite a difficult for the semilocal density functionals because it requires to 
treat both the H-bond (for ice Ih) and strong vdW interactions bond (for $\beta$-AgI) within a great accuracy.   

In Table \ref{tabice} we summarize the performance of the computed lattice constants of both the materials along with 
with the corresponding lattice mismatch from different methods. Note that the PBE functional was studied previously in 
several references. For example, in Ref. \cite{PhysRevB.93.045126}, it was shown that PBE performs better than PBEsol 
because of its accuracy in the determination of the H-bond. 
\begin{table}
\begin{center}
\caption{\label{tabice}Calculated lattice constants (\AA) for ice Ih and $\beta$-AgI as well as the corresponding 
lattice mismatch as obtained with different functionals. Experimental results are taken from 
Refs.~\cite{PhysRevB.87.214101,PhysRevB.93.045126}. The best result for each line is highlighted in bold.}
\begin{ruledtabular}
\begin{tabular}{ccccccc}
 & PBE & SCAN & TM&revTM & Exp. \\
\hline
$a$ (ice Ih) &4.43&4.37&\textbf{4.45}&\textbf{4.45}  & 4.50 \\
$b$ ($\beta$-AgI) &4.69&4.63&\textbf{4.61}&\textbf{4.61}  & 4.59 \\
mismatch &5.7\%&5.8\% &\textbf{3.5\%}&\textbf{3.5\%} & 2.2\% \\
\end{tabular}
\end{ruledtabular} 
\end{center}
\end{table}
In Ref. \cite{PhysRevB.93.045126}, it was also shown that the SG$4$ performs better than PBE and PBEsol. Several 
meta-GGAs like revTPSS and MGGA-MS are showing improved performance because of their ability to describe the 
H-bond~\cite{PhysRevB.87.214101}. However, none of these functional are accurate for both the ice Ih and $\beta$-AgI 
simultaneously. From Table \ref{tabice}, we observe that a better balance in the performance is observed for both the 
ice Ih and $\beta$-AgI with the TM and revTM functional. Both perform considerably well and yield a lattice mismatch 
of 3.5\% which is only a 1.3\% larger than the experimental values. The SCAN functional performs similarly as the PBE 
functional with the mismatch of 5.8\%. The SCAN functional showing underestimation in the lattice constants of ice 
Ih. This is probably due to a bit much short or intermediate-range vdW interaction. Note that SCAN, TM, and revTM 
perform well for the $\beta$-AgI. But TM and revTM are more accurate for ice Ih because of their ability to describe 
the H-bond more accurately. Finally, from the results, we can say that the revTM is a potential candidate for the ice 
lattice mismatch problem.

\subsection{Merit of performance of TM and revTM functionals}

At this point, an elaborating comparison in the performance of the TM and revTM functional from the point of their construction is required. 
Note that the TM functional includes short or intermediate-range vdW interactions in its exchange due to the oscillation of the exchange 
enhancement factor originating from the exchange hole model~\cite{PhysRevLett.117.073001,doi:10.1063/1.4971853,Tang_2018,
doi:10.1021/acs.jpca.9b02921}. However, the revTM functional scales down this short or intermediate-range vdW interactions vdW effects 
of the TM functional by using a slightly different form of the reduced Laplacian gradient in its exchange functional form
~\cite{doi:10.1021/acs.jpca.9b02921}. For the general-purpose solids and molecules, this change does not seem to be great 
consequences except for the dispersion and H-bonded systems for which the short or intermediate-range vdW interaction plays 
a significant role without including any long-range vdW part. Therefore, we can argue from the present study that, due to this 
short or intermediate-range vdW interactions, TM functional shows a bit of overestimation tendency in the behavior of the water 
isomer and cluster which is modified by its revised version i.e, in revTM functional.

For the ice the tendency is quite opposite. In this case inclusion of short or intermediate range vdW interaction actually 
shriken the volume more. In this sense TM functional shows underestimation in volume which is greatly modified by the 
change in the exchange enhancement factor of the revTM functional.

\section{Conclusions}

We have studied the performance of TM and revTM functional along with PBE and SCAN  functionals for the relative 
binding energies of the different water hexamer, and dimers. From the present study, we observe that both the SCAN, TM and revTM 
functional correctly predict the ordering of the stability for different water isomer and cluster. However, the revTM binding energies agree 
quite well corresponding to the MP2 level theory. Regrading the relative stability of the different water hexamer corresponding to 
the most state prism structure, the revTM obtains a very accurate result. For neutral, positively, and negatively charged water 
cluster also, the revTM functional achieves excellent accuracy compared to the PBE, SCAN and TM functionals indicating its 
accuracy to predict the H-bond in the water cluster.  In particular, for water dimer with a single H bond, revTM is consistently 
giving better absolute binding energies over SCAN. Even for the relative energies for hexamer, the revTM is on an average better 
than SCAN. This also implies that revTM predicting the short and intermediate-range interactions quite accurately for water. Considering 
the structural properties of the water monomer, the revTM bit overestimates the results compared to the SCAN functional. Whereas for 
vibrational frequencies the revTM performs with better accuracy for frequency $\nu<1000$~cm$^{-1}$. For different ice phases also 
the revTM shows consistent improvement over the PBE, SCAN and TM functional. It quite accurately predicts the energies and volume of the 
different ice phases. Also, for the ice lattice mismatch problem the TM and revTM functionals show considerable improved performance 
over the PBE and SCAn functionals. 

In conclusion, the productive power of revTM xc functional showing its great accuracy for different problems which may enables to perform several ab-initio calculations 
for water useful for physics, chemistry, biological sciences and material physics.


\section{Simulation details} We performed calculations of the energetic, structural and vibrational properties of 
the water isomer and cluster (TABLE II,  TABLE III, TABLE IV, TABLE V, TABLE VI, TABLE VII, and TABLE VIII) 
in the developed version of the Q-CHEM code~\cite{doi:10.1080/00268976.2014.952696} where 
the XC integrals are calculated with 99 points radial grid and 590 points angular Lebedev grid. The revTM functional results 
are compared with the PBE, SCAN, and TM functional. The basis set used for the respective test set are mentioned in headers of each Table. 

Details of the HB6, DI6, and CT7 (TABLE I) are collected from previous calculation of ref.~\cite{doi:10.1021/acs.jpca.9b02921}, 
where the 6-311++G(3df,3pd) basis set with NWCHEM code with the code recemmended fine grid are used.   

The simulation of the ice phases of water (TABLE IX, TABLE X, and TABLE XI) are calculated using the VASP code~\cite{PhysRevB.47.558} 
with the hardest projector-augmented wave (PAW) pseudopotentials ($H_h$ and $O_h$). The 1200 eV energy cutoff and $4 \times 4 \times 4$ 
$\bf{k}$ mess is used for ice phases. For water we used a simulation box of volume $20 \times 21 \times 22$ \AA$^3$. We optimize the 
ice phases using the respective functional to obtain the energies and volume. The calculations of the $\beta-$AgI is performed in VASP 
with the 1200 eV energy cutoff and $16 \times 16 \times 16$ $\bf{k}$ $\Gamma-$centered $\bf{k}$ mess.     

\section{Acknowledgement}

S.J. would like to thank Dr. Biswajit Santra for providing many useful technical informations about the calculation and manuscript.

\twocolumngrid
\bibliography{reference}
\bibliographystyle{apsrev4-1}

\end{document}